# Proving "new physics" by measuring cosmic ray fluxes

A.Chilingarian, G.Hovsepyan

**Yerevan Physics Institute**

**Abstract**

For more than 100 years, people have measured radiation on the earth's surface, first with the simplest electroscopes and now with sophisticated thousands-ton LHC detectors. With widely available cheap particle detectors, many publications appear on the changes in detector count rate, with attempts to relate them to astronomical objects and events. The authors of these papers don't care about measurement errors, about the influence of meteorological parameters and disturbances of electrical and geomagnetic fields on the fluxes of cosmic rays. They publish "unique" observations during solar and lunar eclipses, Venus's transit over Sun, and comet appearance, during the transit of the Sun across Constellations Libra, Virgo, and many others.

In searches for "new physics," along with multiyear efforts in collecting data, complicated multivariate "graphical" cuts are usually superimposed to "purify" the data and "emphasis" new physics. The natural influence of multiple cuts on the distribution function of the chance probability is usually ignored, and the "optimized" z-score value is applied to standard Gaussian distribution leading to minimal values of chance probability to reject the $H_0$ hypothesis erroneously. We demonstrate that this "naïve" approach leads to erroneous physical inference. We suggest using a distribution function based not only on the z-score value but also on the number of attempts (cuts) applied to achieve a large number of $N\sigma$ (standard deviations from the mean value). Examples of applying the new criteria to experimental data and "toy problems" are discussed.

The paper aims to demonstrate how the measurements of different species of cosmic ray flux can lead to a meaningful physical inference. We want to demonstrate when and how it is possible to path the way from measurement to physical inference and how we can prove that measurements are not artifacts or equipment failures but manifestations of a new physical phenomenon.

1. **Proof in physics**

Physics is an inductive discipline that accepts some assumptions, gathers empirical results, compares them with other experiments and theories, and comes to new inferences.

Standard dictionaries' definitions of "proof":

• The pieces of evidence that compel the mind to accept an assertion as true.

• Argument establishing a fact or the truth of a statement.

- The process of establishing the validity of a statement, especially by derivation from other statements following principles of reasoning.

Thus, scientific proof in physics and other experimental disciplines, by definition, is based on empirical pieces of evidence used to ultimately convince the community that the measurement and inference from it is correct within rigorously defined limits. Well-defined procedures exist for presenting, discussing, confirming, and validating the statements, inferences, and theories based on measurements. In the following sections, we will demonstrate these procedures using as an example measurement of the cosmic ray fluxes on the earth's surface. A final goal is to prove that the peaks in the measured time series are not connected with abrupt changes of atmospheric pressure or outside temperature, are not a failure of equipment or fluctuations due to the finite accuracy of the detector but are a new physical phenomenon and can serve as a basis for the following experimentation for revealing the model and theory of their origination.

For the clarity of the procedure, let's formalize the problem on hand: we are looking for some extraordinary event in the terrestrial atmosphere or in space that influenced the cosmic ray (CR) count rate we measure with particle detectors on the earth's surface. From the registered fluxes, we have to decide if the count rate changes (depletion or enhancement) are due to a new physical effect under question or can be explained by random fluctuations or well-known processes, for instance, abrupt change of the atmospheric pressure.

We have to carefully investigate all dependencies of the count rate, make necessary corrections, and estimate the detector response to different particles. When reporting new physics coming from your measurements, you should return again and again to all possible sources of the experimental errors to prove your inference.

**Particle detectors**

There are plenty of particle detectors registering different species of cosmic rays. Our goal is not to give a review of them or make a comparison analysis or discuss the particle interactions with measuring media. We want to explain the methodology of properly using acquired data for physical inference, i.e., to prove that we measure a genuine signal from a new physical process. Considered examples of the physical inference will be based on measurements made at Aragats cosmic ray (CR) observatory [1]; we will restrict ourselves to minimal information on the detector's operation and refer to data that can be easily downloaded in graphical and numerical formats from the database of cosmic ray division (CRD) of Yerevan Physics Institute (YerPhI)[2].

In Fig. 1, we show some particle detectors operated on Aragats. The primary sensor of the SEVAN network (Fig. 1a, see Chilingarian et al., 2018) consists of standard slabs of 50 x 50 x 5cm$^3$ plastic scintillators. Between two identical assemblies of 100 x 100 x 5 cm3 scintillators (four standard slabs) are located two 100 x 100 x 5 cm$^3$ lead absorbers and a thick 50 x 50 x 25 cm$^3$ scintillator stack (5 standard slabs). Lights capture cones and photomultipliers (PMTs) are located on the top, bottom, and intermediate layers of the detector. In the upper 5 cm thick scintillator, charged particles are effectively registered; however, for the registration of neutral particles, there needs to be more substance. When a neutral particle traverses the top thin (5cm) scintillator, usually no signal is produced. Incoming neutral particles undergo nuclear reactions in the thick 25 cm plastic scintillator and produce charged particles. The absence of the signal in the upper scintillators, coinciding with the signal in the middle scintillator, indicates neutral particle traversal (gamma-ray or neutron). Data Acquisition (DAQ) electronics provide registration and storage of all logical combinations of the detector signals for further offline analysis and for online alerts issuing, thus. If we denote by "1" the signal from a scintillator and by "0" the absence of a signal, then the following combinations of the 3-layered detector output are possible: "111" and "101"—traversal of high energy muon; "010"—traversal of a neutral particle; "100"—traversal of low energy charged particle stopped in the scintillator or the first lead absorber (energy less than ≈100 MeV). "110"—traversal of a charged particle of higher energy, which stopped in the second lead absorber. "001"—registration of inclined charged particles. The Data Acquisition electronics (DAQ) allows the remote control of the PMT high voltage and other detector parameters. The total weight of the SEVAN detector, including steel frame and detector housings, is ≈1,5 tons. 10 SEVAN detectors operate in Armenia, countries in Eastern Europe, and Germany.

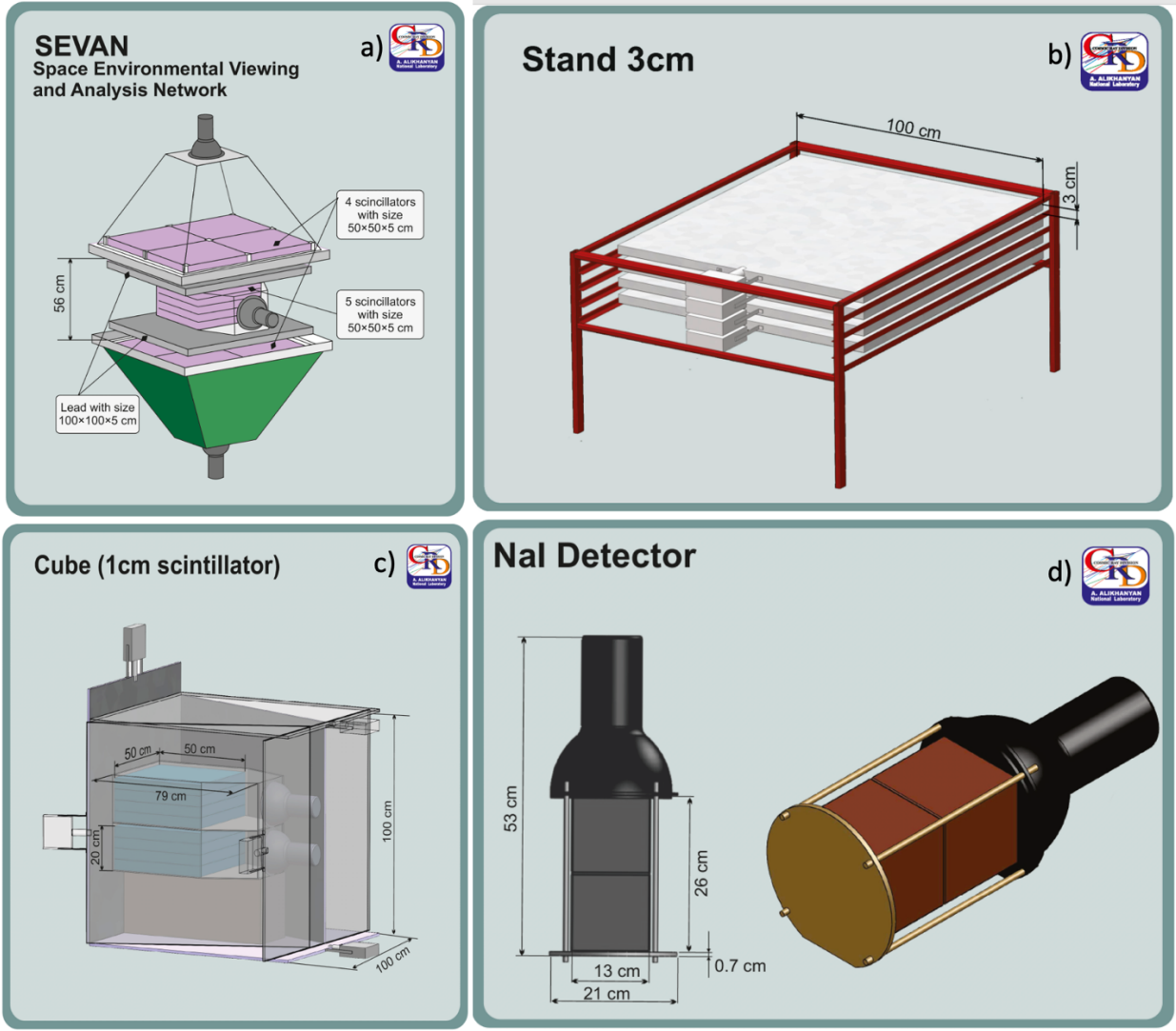

**Figure 1. Particle detectors operated on Aragats**

The "STAND3" detector comprises four layers of 1-cm-thick, 1-m$^2$ sensitive area scintillators stacked vertically see Fig. 1b. The High Energy Physics Institute, Serpukhov, Russian Federation, fabricated these scintillators. The light from the scintillator through optical spectrum-shifter fibers is reradiated to the long-wavelength region and passed to the photomultiplier (PMT FEU-115M). The maximum luminescence is emitted at the 420-nm wavelength, with a luminescence time of about 2.3 ns. The STAND3 detector is tuned by changing the high voltage applied to the PMT and by setting the thresholds for the shaper-discriminator. The discrimination level is chosen to guarantee both high efficiency of signal detection and maximal suppression of photomultiplier noise. Coincidences of the signals from 4 layers allow selecting charged particles with energy thresholds from 10 MeV ("1000" coincidence, the signal only in the upper layer) to 60 MeV ("1111" coincidence, signals in all layers)

The Cube assembly (Fig. 1c) consists of two 20-cm thick scintillators of 0.25-m$^2$ area each, enfolded by 1 cm thick, 1-m$^2$ area scintillators. This design ensures that no charged particle may hit the inside 20 cm without hitting the surrounding "veto" scintillators. The 20-cm thick plastic scintillators are overviewed by the photomultiplier PM-49 with a large cathode operating in a low-noise regime. Surrounding detectors (six units) are 1-cm thick molded plastic scintillators. The efficiency of registration of neutral particles by 1-cm thick scintillators is 1-2% and weakly depends on their energy. The energy losses of passing electrons and muons in a 20-cm-thick plastic scintillator are ~40 MeV. Taking into account the construction material of the detector (2-mm iron tilt and 1-cm plastic scintillator) and the roof of the building (1-mm iron tilt), the electron registration energy threshold for the upper 20-cm-thick scintillator is estimated to be about 10 MeV and for the bottom one ~40 MeV. The efficiency of gamma ray registration in a 20 cm thick scintillator equals ~20%, and the neutron detection efficiency is ~27%.

The NaI detector network measuring particle energy consists of 4 Na( (Tl) spectrometers (Fig. 1d) packed in the sealed 3-mm- thick aluminum housing. Each crystal is coated with 0.5 cm of magnesium oxide (MgO) on all sides (because the crystal is hygroscopic) with a transparent window directed to the photo-cathode of an FEU-49 PMT. The large cathode of PMT (15 cm diameter) provides an excellent light collection. The spectral sensitivity range of FEU-49 is 300–850 nm, which covers the spectrum of the light emitted by NaI(Tl). The sensitive area of each NaI crystal is ~0.0348 m$^2$, the total area of the four crystals is ~0.14 m$^2$, and the gamma-ray detection efficiency is 60-80%. A logarithmic analog-digit converter (LADC) is used to code PM signals. Calibration of LADC and code-energy conversion was made by detecting the peak from the 137Cs isotope emitting 662 keV gamma rays and by the muon peak (appeared at ≈ 50 MeV) in the histogram of energy releases in the NaI crystal. The PMT high voltage was tuned to cover both peaks in the histogram of LADC output signals to ensure linearity of LADC in the energy region of 0.3–50 MeV. A significant amount of substance above the sensitive volume of NaI crystals (0.7 mm of roof tilt, 3 mm of aluminum, and 5 mm of MgO) prevents electrons with energy lower than ~3 MeV from entering the sensitive volume of the detector. Thus, the network of NaI spectrometers below 3 MeV can detect gamma rays only.

2. **Detector response function, purity and efficiency of the detector, and detector response to charged and neutral CR species.**

The count rate of any detector depends on its size, geographic location, and registration efficiency. The count rate is influenced by atmospheric pressure, the gradient of outside temperature, the near-surface electric field (NSEF), and the geomagnetic field and solar wind. The count rate depends on oscillations in the power supply lines, in the transformers, the day, and year periodicities, the noise due to the random character of physical processes used for particle detection, etc.
To derive parameters having physical meaning, we have to deconvolute the measured count rate to be not dependent on the specific characteristic of the detector, estimate different particle fluxes and energy spectra as they were before entering the detector, the CR flux and determine errors, without which presenting of any experimentally measured quantity is senseless.

First, we have to investigate the response of the particle detector to different particle fluxes. It is the so-called direct problem of CR: for the given particle fluxes, determine (measure or simulate) the count rates. The best way for it is calibrations with particle beams on man-made accelerators. However, this is not an easily accessible option. Thus, we use CR flux generators (for instance, EXPACS [3], giving flux of all CR species on all latitudes, longitudes, and altitudes) and GEANT4 code [4] (a standard tool for high-energy physics experiments) for the detector response modeling. Thus, additional "model" errors will influence the experimentally measured values and will harm the physical inference from it (see the methodology of making a physical inference based on simulations in [5]). Nonetheless, there is no way to avoid it; without simulations, the measured fluxes are arbitrary, and physical inference is senseless.

In Table 1, we show the so-called purity of the STAND3 detector (Fig. 1b) coincidences o obtained with EXPACS and GEANT4 packages. As we can see from the Table, the "1000" coincidence efficiently selects neutrons (17.2%) and gamma rays (33.6%), "1111" coincidence – muons (77%), and "100" – low energy muons and electrons (80%). Thus, the STAND3 detector can investigate three types of secondary CRs.

**Table 1. Purity of the STAND3 coincidences measuring the ambient population of secondary cosmic ray flux (background) flux on Aragats (3200 m) in percent**

| Second. type | Purity (%) | | | |
|---|---|---|---|---|
| | 1000 | 1100 | 1110 | 1111 |
| n | 17.22 | 5.83 | 2.20 | 0.48 |
| p | 4.37 | 7.42 | 7.28 | 6.23 |
| mu+ | 5.01 | 12.21 | 23.03 | 41.28 |
| mu- | 4.52 | 11.08 | 20.04 | 35.92 |
| e- | 21.31 | 24.77 | 19.95 | 6.85 |
| e+ | 13.44 | 18.62 | 15.67 | 6.67 |
| gamma | 33.57 | 20.07 | 11.83 | 2.57 |

Only purity calculations are not enough to characterize the detector. We also need the efficiency of registration, which usually is a function of particle energy. Thus, in Table 2, we show the energy dependence of the registration efficiency for other coincidences of the STAND3 detector (Fig1b). From the table, we can see that the coincidences of the detector layers strongly depend on the electron energy. "1000" coincidence (signal attenuates after the upper scintillator) selects effectively electrons with energies 10 -20 MeV; "1100" coincidence – with energies (20-30 MeV); "1110" – with energies 30-40 MeV; and "1111" – above 50 MeV.

Table 2. Efficiency(%) of electron registration by STAND3 detector.

| STAND3 | Scin. 1 | Scin. 2 | Scin. 3 | Scin. 4 | 1000 | 1100 | 1110 | 1111 | Sum |
|---|---|---|---|---|---|---|---|---|---|
| 10MeV | 81.27 | 0.19 | 0.16 | 0.14 | 81.23 | 0.01 | 0.00 | 0.00 | 81.25 |
| 12MeV | 89.58 | 0.31 | 0.27 | 0.22 | 89.34 | 0.10 | 0.00 | 0.00 | 89.43 |
| 14MeV | 92.09 | 0.50 | 0.39 | 0.34 | 91.45 | 0.29 | 0.00 | 0.00 | 91.73 |
| 16MeV | 93.12 | 5.72 | 0.52 | 0.45 | 87.28 | 5.22 | 0.00 | 0.00 | 92.50 |
| 18MeV | 93.66 | 23.98 | 0.71 | 0.62 | 69.97 | 22.72 | 0.00 | 0.00 | 92.69 |
| 20MeV | 94.07 | 44.31 | 0.96 | 0.80 | 50.43 | 42.23 | 0.03 | 0.00 | 92.70 |
| 30MeV | 94.78 | 82.54 | 23.79 | 2.13 | 13.08 | 57.41 | 20.73 | 0.03 | 91.26 |
| 40MeV | 95.00 | 89.10 | 63.09 | 14.53 | 6.22 | 25.89 | 47.18 | 10.29 | 89.58 |
| 50MeV | 95.06 | 91.28 | 77.17 | 44.70 | 4.10 | 14.10 | 33.13 | 37.47 | 88.79 |
| 60MeV | 95.21 | 92.38 | 82.99 | 61.91 | 3.08 | 9.41 | 22.86 | 53.19 | 88.54 |

In Table 3, we show the efficiencies of gamma ray registration by the STAND3 detector. In the Table, we can see that although detector layers register high-energy (>20 MeV) gamma rays with an efficiency of 20% and more, the efficiency of coincidences is much lower. Thus, for separating the mixed electron–gamma flux, the usage of STAND3 coincidences for the comparison of simulated and measured count rates is preferable.

Table 3 Efficiency(%) of gamma rays registration by STAND3 detector.

| STAND3 | Scin. 1 | Scin. 2 | Scin. 3 | Scin. 4 | 1000 | 1100 | 1110 | 1111 | Sum |
|---|---|---|---|---|---|---|---|---|---|
| 10MeV | 5.88 | 6.03 | 5.48 | 4.94 | 5.81 | 0.03 | 0.00 | 0.00 | 5.84 |
| 20MeV | 7.07 | 10.77 | 10.13 | 9.15 | 4.55 | 2.35 | 0.02 | 0.00 | 6.92 |
| 30MeV | 7.43 | 13.99 | 14.67 | 13.42 | 2.06 | 4.22 | 0.89 | 0.01 | 7.18 |
| 40MeV | 7.63 | 15.50 | 18.64 | 18.02 | 0.83 | 3.56 | 2.52 | 0.37 | 7.28 |
| 50MeV | 7.90 | 16.29 | 21.25 | 21.89 | 0.44 | 2.37 | 3.16 | 1.50 | 7.47 |
| 60MeV | 8.09 | 16.77 | 22.92 | 25.04 | 0.29 | 1.43 | 3.10 | 2.83 | 7.65 |

3. **The influence of the atmospheric parameters on the particle detector count rates**

For 80 years on Aragats station, continued measurements of the different species of secondary cosmic rays and ultra-high energy primary cosmic rays. Electrons, muons, and gamma rays are measured with NaI and plastic scintillators, and neutrons with neutron monitors and hybrid SEVAN detectors, cores of extensive air showers – with neutron monitors. Energy spectra are recovered with NaI spectrometers and scintillation spectrometers.
Davis Vantage Pro2 weather station includes a rain collector, temperature sensor, humidity sensor, anemometer, solar radiation sensor, ultra-violet (UV) radiation sensor, and others.
The near-surface electrostatic field changes were measured by a network of six field mills (Boltek EFM-100), three of which were placed in Aragats station, one in Nor Amberd station at a distance of 12.8 km from Aragats, in Burakan, 15 km from Aragats, and in Yerevan, 39 km from Aragats. 3 components of the geomagnetic field are measured with LEMI- 018 vector magnetometer. All data are entering the Advanced data extraction infrastructure (ADEI) at CRD/YerPhI providing vast possibilities for multivariate visualization and correlation analysis. In Fig.2, we show atmospheric parameters influencing the count rate of particle detectors. The most important of them are outside temperature (red), atmospheric pressure (magenta), and NSEF (black). The count rate bias due to atmospheric effects can reach 10%, surpassing all subtle effects expected from astrophysical sources. Thus, special corrections should be made to disentangle possible "new physics" from the simple biases of the atmospheric nature.

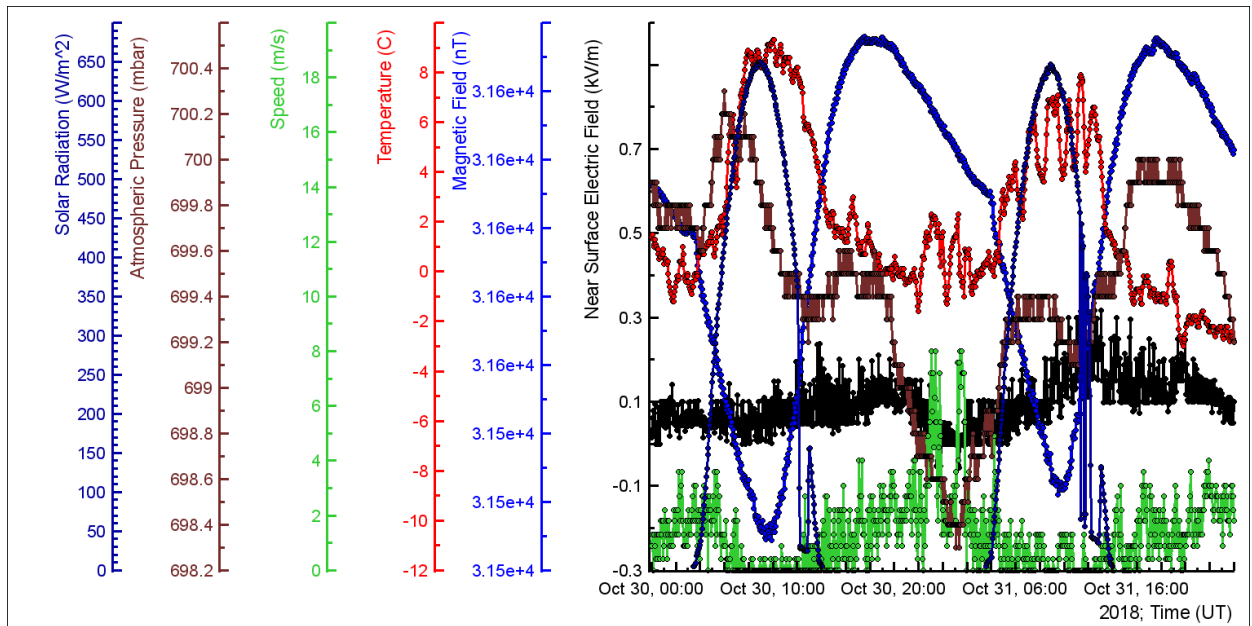

**Figure 2. Time series of environmental parameters influencing particle detector count rates. Black – NSEF; Blue - geomagnetic field; red - outside temperature; green – wind speed, magenta – atmospheric pressure.**

In Fig. 3 we demonstrate the influence of mentioned parameters on the count rates of NaI detectors (blue curve).

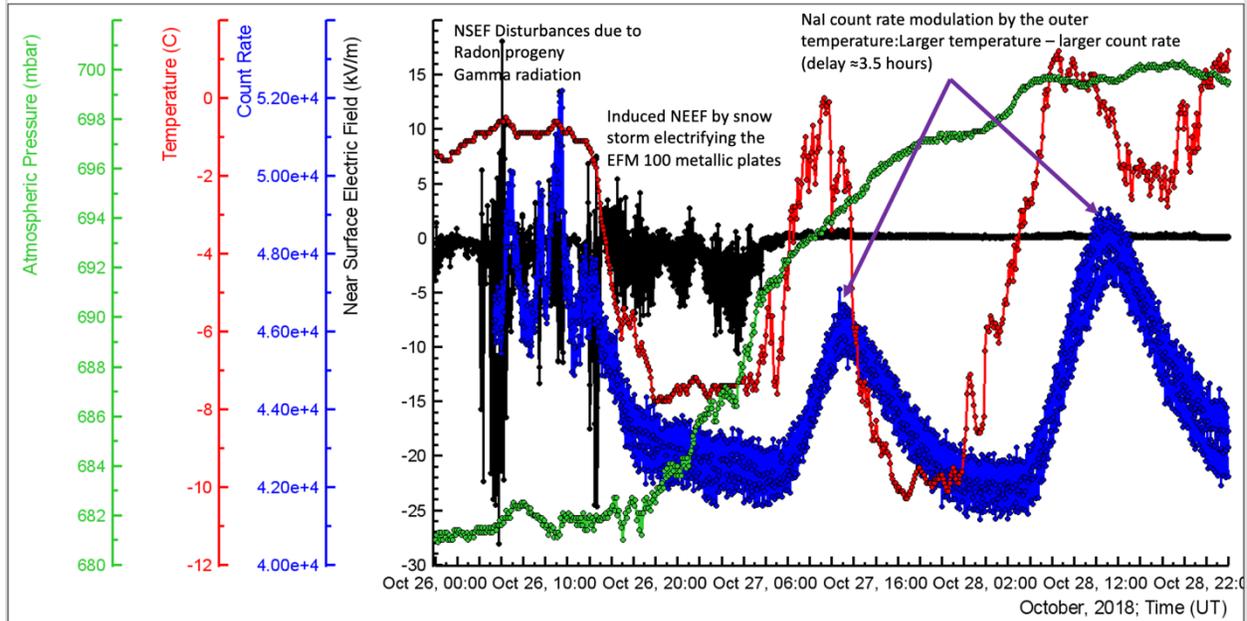

**Figure 3.** 1-minute time series of NaI detector count rates (blue), influenced by NSEF disturbances (black) and outside temperature (red). The influence of temperature on the count rate is overwhelming, and the influence of the large variations of atmospheric pressure (green) is not noticed.

In Fig. 4, we show an example of correlation analysis of time series. The NaI crystals are located under the metallic roof of the SKL experimental hall on Aragats. During sunny days (solar irradiation is shown in red), the roof transfer heat to detectors (although with a delay of 3.5 hours, see green vertical lines in the picture and inset), significantly increasing the count rate (black curve, 12% increase).

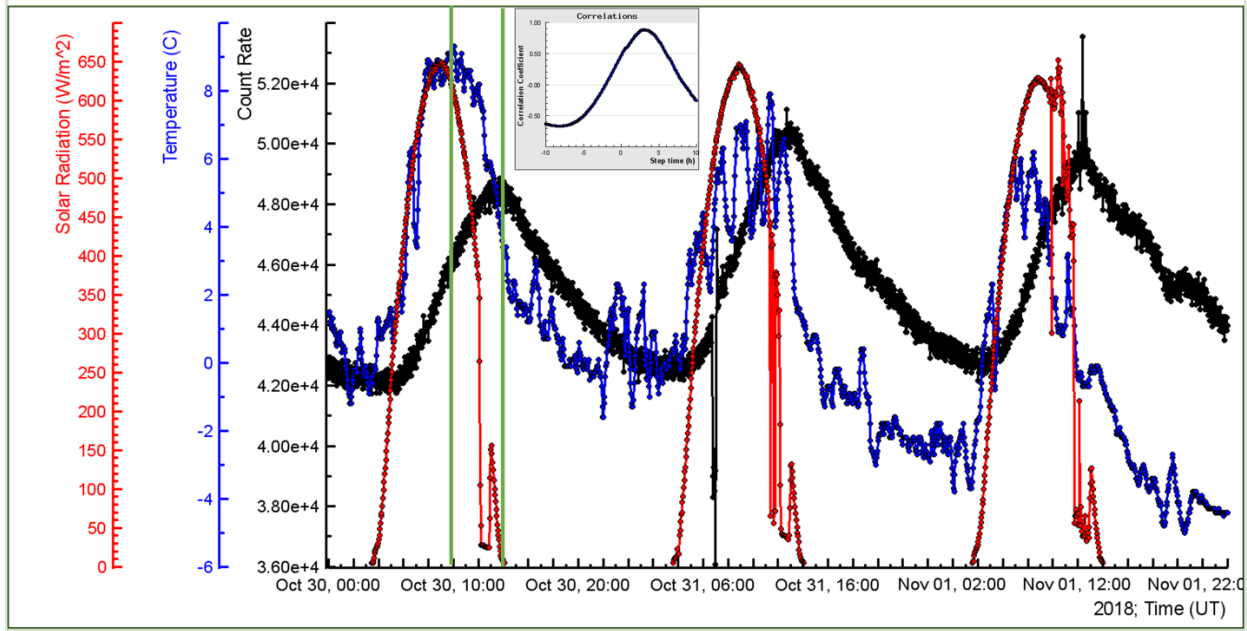

Figure 4. 1mute count rate of NaI detector (black curve) influenced by outside temperature variations (blue curve). By the red curve, solar irradiation is shown. In the inset, we show the delayed correlation curve.

The count rate of scintillation detectors is anti-correlated with the temperature and is most influenced by atmospheric pressure. There are several well-known reasons for the count rate variations, and all of them should be carefully examined before discussing a physical inference based on the CR measurements.

4. **Statistical moments of measured count rates, relative errors, and significance of detected peaks. Gaussian nature of random errors.**

For the investigation of the detector parameters, we should choose a period corresponding to more or less stable weather conditions that do not seriously influence the count rate of the detector. Count rates are characterized by the statistical moments; the sampling estimates of these parameters are means and variances. After measuring the count rates with inherent fluctuations, we have to decide if the measured variation (enhancement or depletion) is within acceptable limits or if it is an extraordinarily outburst manifesting new physics (or detector failure). We will show the technique of finding the genuine peaks taking as an example the recently observed thunderstorm ground enhancements (TGEs, [6]) on Aragats.

Summer 2022 in Aragats was dry and hot. The particle flux enhancement during very few "summer TGEs" never exceeds 8%, and the corresponding peak significance measured in the number of standard deviations above fair-weather value never exceeds 10. And suddenly, on September 22, during an ordinary storm, detectors registered 7 TGEs, 3 of which with very large flux enhancement, see Fig 5, where four coincidences of STAND3 detector are depicted. For digitizing and comparing flux enhancements, we calculate means and variances on fair weather when all meteorological parameters were stable (see Figure 6).

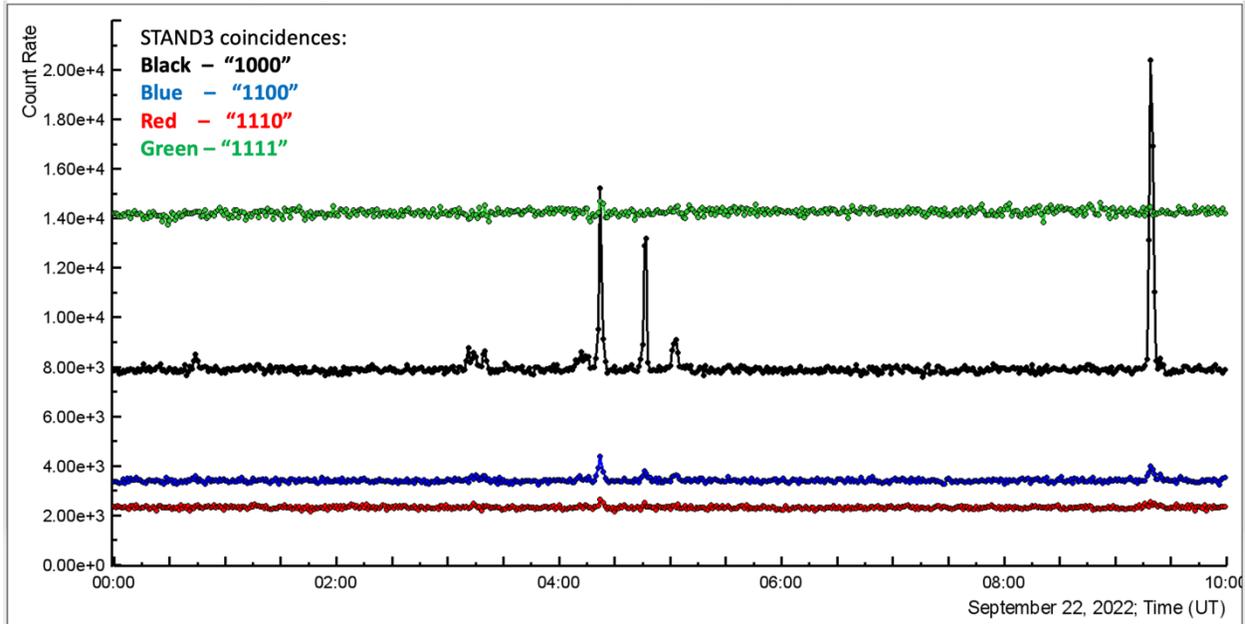

**Figure 5.** 1-minute time series of the count rates of STAND3 detector's coincidences

The presented in Fig. 6 distributions, their means, variances, and relative errors give a measure of sensitivity to the "new physics"; the relative error outlines the minimum signal value, which can be considered a possible artifact. The relative errors of coincidences are slightly different; however, a limit of 3% can be accepted as a conservative estimate. All measured fluctuations within 3% cannot be accepted as a significant deviation from the mean value to be examined for possible nontrivial signals. In Fig. 7, we show time series of count rates for one of 7 TGE events, in which we see peaks in all STAND3 detector coincidences. The number of standard deviations (critical value, N$\sigma$) for each peak is calculated using data from Fig.6. According to Neumann – Pearson's approach to statistical decisions [7], a critical value is fixed and used to accept or reject the so-called $H_0$ hypothesis that all events (including giant outburst) belong, for instance, to the Gaussian population (the process in control). Each critical value (usually set to 3 in medical research and 5 in elementary particle searches, see, for instance, Fig.8 in [8],[9]) is connected to the so-called p-value, the integral of the Gaussian function from the critical value to infinity. To prove the existence of a signal of "new physics," we have to reject $H_0$ with the maximal possible confidence. However, significant deviations from $H_0$, i.e., a very low probability of $H_0$ being true, do not imply that the opposite hypothesis is automatically valid. As was mentioned by Astone and D'Agostini [10], behind the logic of standard hypothesis testing is hidden a revised version of the classical proof by contradiction. "In standard dialectics, one assumes a hypothesis to be true, then looks for a logical consequence manifestly false, to reject this hypothesis. The "slight" difference introduced in the statistical tests is that a "very improbable consequence" replaced the false consequence.

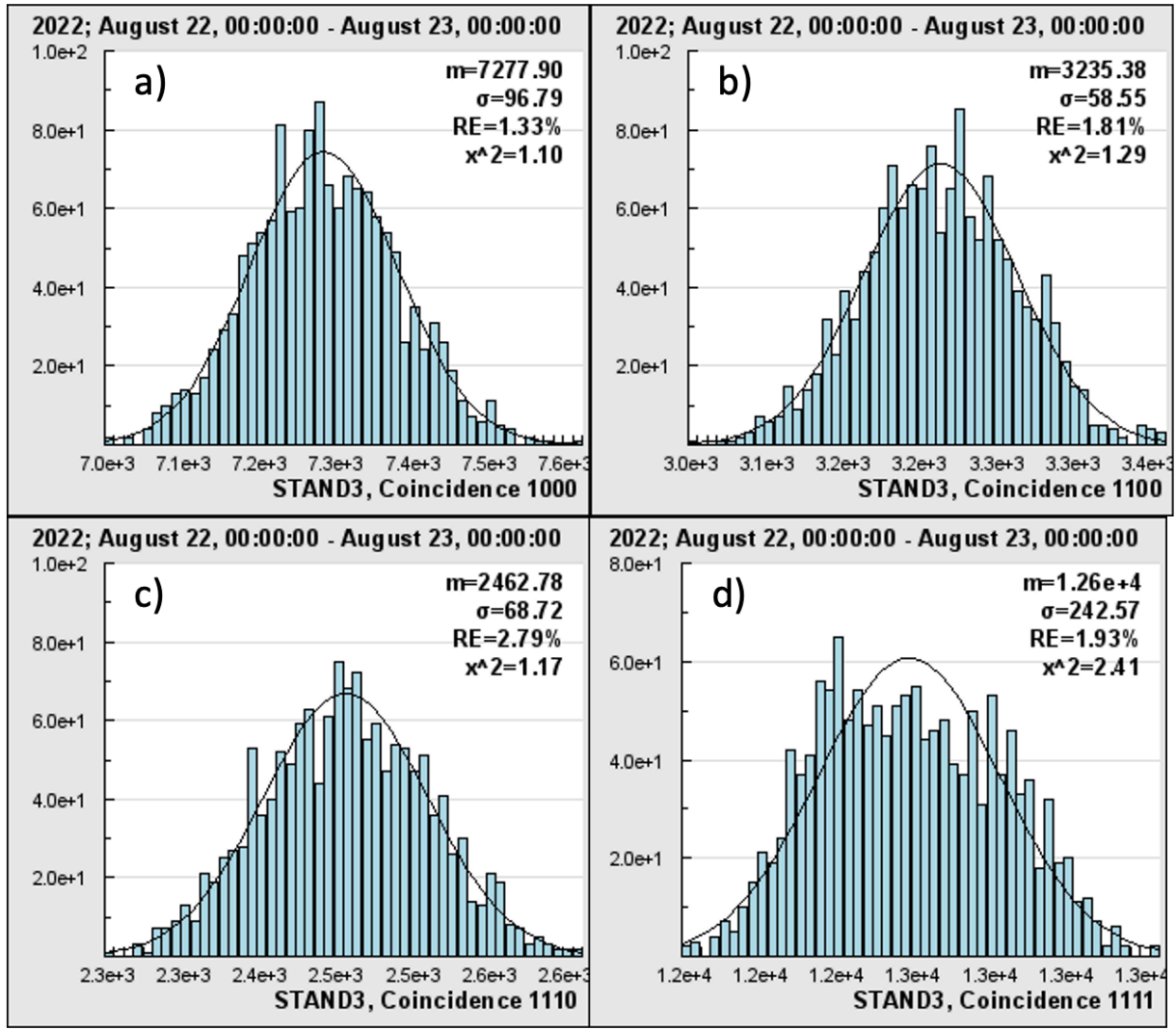

*Figure 6.* **Means, variances, and relative errors of STAND3 detector coincidences. The whole day August 22, 2022, 1440 minutes (no corrections to the atmospheric pressure and outside temperature are done)**

In Table 4, we compare the simulated and measured 1-minute count rates of STAND3 coincidences. The flux discrepancies are within 20%, which is satisfactory for the approximate background fluxes obtained from the WEB calculator [3] and for integrating the particle flux for a whole day, neglecting the so-called day-wave, the variation of the flux during a day due to changing meteorological conditions.

**Table 4. Simulated with EXPACS and GEANT4 and measured 1-minute count rates of STAND3 coincidences**

| STAND3/min  | "1000" | "1100" | "1110" | "1111" |
|---|---|---|---|---|
| Simulation  | 7278   | 3235   | 2197   | 15328  |
| Measurement | 8617   | 3464   | 2463   | 12600  |

For proving the TGE at 04:22 – 04:23, we do not fix the critical value but calculate it according to the measured peak height, see Fig.7. The critical value is much larger than 5, and the corresponding p-value (chance probability to erroneously reject the $H_0$) is extremely small. In Fig.9a, we show the chance probability corresponding to the critical value of 5 (there is only one chance from 3500000 to reject $H_0$ erroneously). The Gaussian integral from 82 (the peak significance in the "1000" time series) to infinity equals the enormously small number of $3.8 \times 10^{-1461}$, see Fig. 9b. It will be difficult to formulate any hypothesis that will be more improbable than $H_0$ (that signal is fluctuation only) for this chance probability. Our concept of reality breaks against numbers like that. And that isn't even the smallest number out there; we have measured a TGE significance equal to $100\sigma$ and even more.

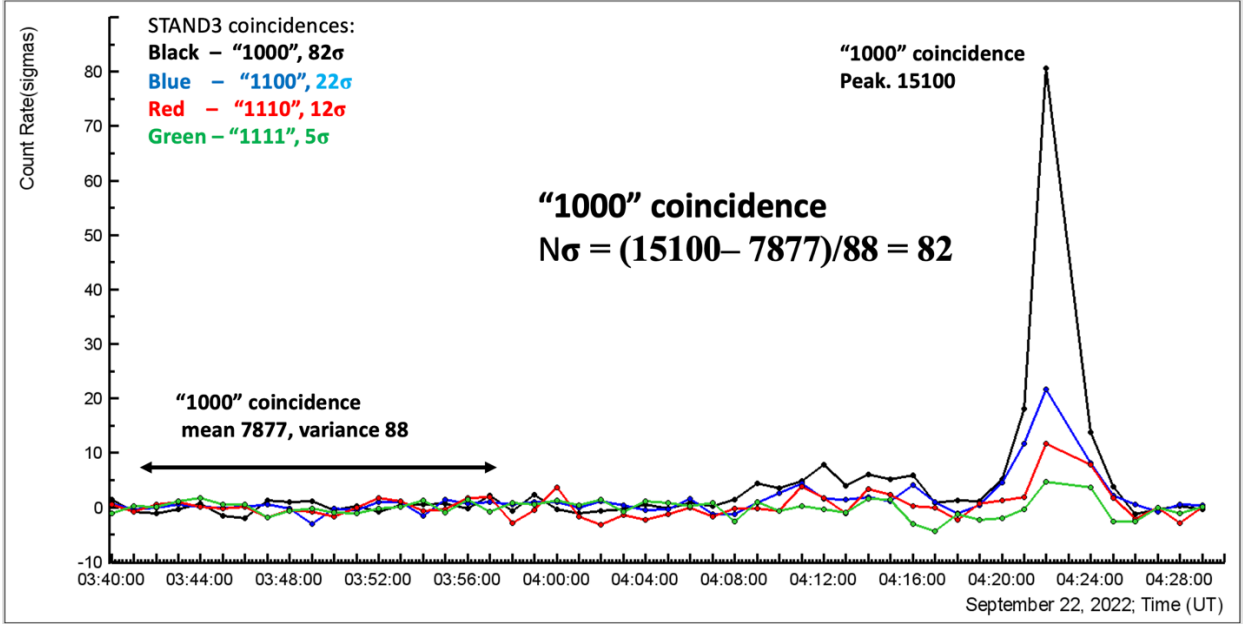

**Figure 7. 1-minute time series of the count rates of STAND3 detector coincidences in numbers of standard deviations, critical value, Nσ.**

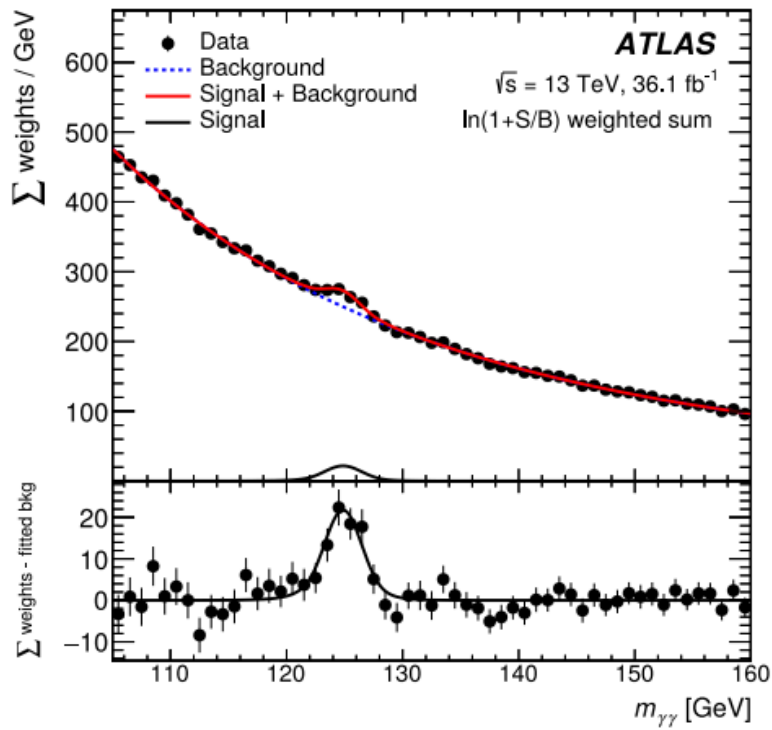

**Figure 8.** The distribution of the invariant mass of the two photons in the ATLAS experiment at LHC. Measurement of H→γγ using the full 2015+2016 data set. An excess is observed for a mass of ~125 GeV. In the bottom panel – the background subtracted distributions. Adopted from Biglietti and all. 2022, *J. Phys.: Conf. Ser.* **1586** 012028

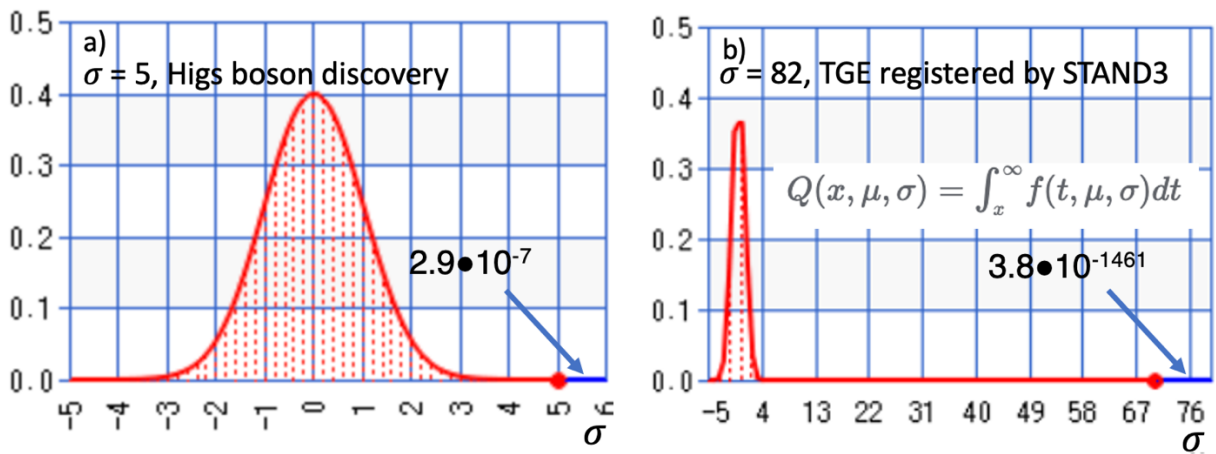

**Figure 9.** Standard Gaussian distribution demonstrating the chance probability (blue lines) of Higgs boson evidence (a), and detection of TGE by STAND3 upper scintillator (b). Chance probability was calculated by the ke!san Online Calculator (https://keisan.casio.com/)

## 5. Observer's influence on the distribution function of measurements

Usually, experimental physicists superimpose different selection criteria on the experimental dataset to obtain a subsample in which outstanding measurements significantly differ from the general population. It is a common practice, and we can refer to searches of celestial objects responsible for ultra-high energy cosmic rays, solar protons emitted during flares, maximums in the invariant mass distribution (cuts in the phase space), etc.
The final goal of such an optimization is obtaining the most significant "sigma" (z-score)! Physicists forget that the z-score value has no statistical meaning if the underlying distribution function is not specified; only chance probability has mining, i.e., the probability of erroneously rejecting the $H_0$ hypothesis.

Performing N measurements, we assume that random variables in the obtained sample are IID: independent and identically distributed. And we are looking for the measurements that significantly differ from the mean value (a Gaussian distribution is assumed). In each sample, we can find an extreme value (maximum or minimum), and it is convenient to present the extreme value using the z- score (normalized N (0,1) distribution). Then we calculate the number of standard deviations from the mean value obtained under the $H_0$ - N$\sigma$. For the N(0,1) distribution, the chance probability of erroneously rejecting $H_0$ (p-value) for the right-tailed test equals:

$$G^{>N} = \int_{N}^{\infty} e^{-\frac{x^2}{2}} dx \quad (1)$$

Thus, applying M different selections (cuts) in the attempt to obtain the maximum z-score and, consequently, minimal p-value, we get a sample of p-values corresponding to the maximum z-score of each of M cuts,

$$G^{>N}_i, i=1, M. \quad (2)$$

The distribution function of the maximum value from M calculated, $N_{max}$ can be readily obtained from the binomial distribution.

$$GE^{>Nmax} = M \star G^{>Nmax} (1 - G^{>Nmax})^{M-1}, \quad (3)$$

After making M selections, the physicist selects the larger $N_{max}$ and usually publishes the inference that the celestial source [13,14], penta-quark[15], etc., are observed with a chance probability calculated with equation (1) with a crucial difference that $N_{max}$ is used instead of the initial N!
But the distribution of $N_{max}$ is not equivalent to the distribution of initial measurements, and using equation (1) instead of equation (3) for calculating p-value can lead to obtaining the chance probability of 2-4 orders of magnitude less than using equation (3). As an example of the erroneous calculation of the p-value below, we describe a paper aiming to "improve" the detection of high-energy protons by the L3 detector during the July 14, 2000, intense solar flare and ground-level enhancement (Bastilian GLE). The initial z-score was published as 4.2[16], and the obtained after 4100 tests – was 5.7 [17].

Protons accelerated nearby the sun during energetic solar events could sometimes unleash large particle showers in the terrestrial atmosphere and initiate additional fluxes of particles detected on the earth's surface. The CERN-based L3+C detector system combined the high-precision muon drift chambers of the L3 spectrometer with an air shower array on the surface. The detector was located near Geneva (6.02°E, 46.25°N) at an altitude of 450 m above sea level and about 30 m underground, providing an average energy threshold of around 20 GeV for vertically incident muons. The full geometrical acceptance was ~200 m$^2$ sr, covering a zenith angle ranging from 0° to 60°. The muon drift-chamber system installed in a 1000 m$^3$ magnetic field of 0.5 Tesla was used to record cosmic ray muons and to measure their momenta precisely.

All selected events were binned according to a specific live-time interval and muon's arrival directions on the ground. The L3+C data-taking system used a live-time interval of 0.839 s as a minimal time bin in counting the number of events within this interval. 100 such bins were combined to form an 83.9 s live-time bin as the primary time unit in searching for possible signals. The direction cosines l = sinθ cosφ and m = sinθ sinφ were used as measurables of the muon directions, where θ and φ are the zenith and azimuth angles of the muon direction at the surface. The squared area of the variables l and m was divided into a 10 x10 (l, m) grid. Ignoring those cells with poor statistics within the detector acceptance, 41 sky cells remained for the investigation. The contour lines for directions having an equal event rate are shown in Fig. 1 of [17]. The excess appeared at a time just coincident with the peak increase of lower energy solar protons and after the X-ray flare started. The background distribution was measured in the same sky cell 12 h before 10:00 UT with 18.18 min live-time bins. Using the fitted mean of 255 and the standard deviation equal to 13.7, the excess of 78 events gives rise to a 5.7σ effect.

Thus, after 42*100 tests, the p-value, according to 5.7σ from equation (1), corresponds to ≈ 6*10$^{-9}$ chance probability; however, using the correct equation (3) – it equals 2.5*10$^{-4}$, equivalent to critical value N$_{max}$ of less than 4σ, i.e., less than initially reported 4.2σ.

## 6. Conclusions

After considering all possible influences of atmospheric parameters, electronics or power outages, and random fluctuation on the particle detector count rate becomes possible to prove the statement's validity on the existence of a significant enhancement of the count rate connected with the new physical phenomenon. This inference is based on the careful estimation of the detector response function, comparison of the count rates of the different detectors at the same location, monitoring of the atmospheric conditions, and calculations of the chance probability of possible erroneous decisions (Fig. 9). In the last section we demonstrated how the cuts superimposed on the initial data can artificially lower the p-value and lead to erroneous enlargement of the reported physical result significance.

The next steps in establishing the new physical phenomenon are connected with revealing the origin of the new phenomenon, including measurement of energy spectra of TGE electrons and gamma rays, performing simulations of the particle propagation in the atmosphere, and comparing simulation and experimental data. At each step, well-established procedures ensure the correctness and soundness of the physical inference. The ASNT spectrometer can measure the electron and gamma ray energy spectra separately. Using these spectra, with GEANT4 code, we calculate the expected (modeled) count rates of TGE electrons and gamma rays and compare them with experimentally measured ones. At Aragats cosmic ray observatory, various particle detectors are monitoring CR fluxes and energy spectra simultaneously, giving the possibility of cross-calibration.

An exhausting demonstration of the techniques of physical inference can be found in references [9,11,12], containing proofs of the existence of the Higgs boson, the gravitational waves, and the signal from the CRAB nebula detected by the LHAASO experiment.

**Acknowledgment**

Authors thank A.Kavalov and S.Chilingaryan for valuable discussions.

**References**

1. A. Chilingarian, K. Avagyan, V. Babayan et. al., Aragats space-environmental center: Status and SEP forecasting possibilities, J. Phys. G 29, 939 (2003).
2. CRD portal:http://adei.crd.yerphi.am/adei/
3. T. Sato, Analytical model for estimating the zenith angle dependence of terrestrial cosmic ray fluxes, PLOS ONE 11 (2016) e0160390.
4. GEANT4 collaboration, GEANT4–a simulation toolkit, Nucl. Instrum. Meth. A 506 (2003) 250.
5. A.Chilingarian, Nonparametric methods of data analysis in cosmic ray astrophysics, an applied theory of Monte Carlo statistical inference 2004, ISTC A757 project, http://crd.yerphi.am/Monogram
6. A. Chilingarian, G. Hovsepyan, D. Aslanyan, T. Karapetyan, Y. Khanikyanc, L.Kozliner, B. Sargsyan, S.Soghomonyan, S.Chilingaryan, D.Pokhsraryan, and M.Zazyan (2022) Thunderstorm Ground Enhancements: Multivariate analysis of 12 years of observations, Physical Review D 106, 082004 (2022).
7. E. L. Lehmann 1993, "The Fisher, Neyman-Pearson theories of testing hypotheses: One theory or two?," J. Amer. Statist. Assoc88, 1242.
8. Biglietti and all. 2022, J. Phys.: Conf. Ser. 1586 012028.
9. S. Manzoni 2019, Physics with Photons Using the ATLAS Run 2 Data, ISSN 2190-5053, Springer Theses, ISBN 978-3-030-24369-2 https://doi.org/10.1007/978-3-030-24370-8
10. Astone, P., D'Agostini, G. Inferring the intensity of Poison processes at the limit of the detector sensitivity, CERN-EP/99-126, 1999.Proof_in+Physic
11. B. P. Abbott, et al. 2020, A guide to LIGO–Virgo detector noise and extraction of transient gravitational-wave signals, Class. Quantum Grav. 37 055002.
12. Nie L, Liu Y, Jiang Z, and Geng X. Ultra-high-energy Gamma-Ray Radiation from the Crab Pulsar Wind Nebula, 2022. The Astrophysical Journal 924 42

13. Abraham et al., (2007), Correlation of the highest-energy cosmic rays with the positions of nearby active galactic nuclei, Astroparticle Physics 29 (2008) 188–204.
14. A.Chilingarian, G.Karapetyan, et al., Statistical Methods for Signal Estimation of Point Sources of Cosmic Rays, (2006) Astroparticle Physics, Astroparticle Physics, 25, pp. 269-27615.
15, C. Seife, Rara Avis, or Statistical Mirage? Pentaquark Remains at Large, Science, 306 (2004) 1281.